\documentclass[%
 reprint,
superscriptaddress,
 amsmath,amssymb,
 aps,
]{revtex4-2}

\usepackage{graphicx}
\usepackage{dcolumn}
\usepackage{bm}
\usepackage{algorithm2e}

\usepackage{hyperref}

\usepackage{flushend}
\usepackage{todonotes}

\begin{document}

\title{Ultra-Large-Scale Compilation and Manipulation of Quantum Circuits with Pandora}

\author{Ioana Moflic}
\email{ioana.moflic@aalto.fi}
\affiliation{Aalto University, Espoo, Finland}

\author{Alexandru Paler}
\email{alexandru.paler@aalto.fi}
\affiliation{Aalto University, Espoo, Finland}

\begin{abstract}
There is an enormous gap between what quantum circuit sizes can be compiled and manipulated with the current generation of quantum software and the sizes required by practical applications such as quantum chemistry or Shor's algorithm. We present Pandora, an efficient, open-source, multithreaded, high-performance-computing-enabled tool based on circuit rewrites. Pandora can be used for quantum circuit equivalence checking, full compilations of large circuits, and scalable, streaming quantum resource estimation frameworks. Pandora can easily handle billions of gates and can stream circuit partitions in resource estimation pipelines at very high rates. We utilized Pandora for full compilations of Fermi-Hubbard 100x100 and 1024-bit Shor's algorithm circuits. Compared to TKET and Qiskit, we determine a performance advantage for manipulating circuits of more than 10000 gates. For equivalence checking tasks, Pandora outperforms MQT.QCEC on specific circuits that have more than 32 qubits. The performance and versatility of Pandora open novel paths in quantum software.
\end{abstract}

\maketitle

\section{Introduction}

Quantum software for compilation and manipulation of quantum circuits (i.e. quantum circuit design automation) needs to reach the scale of classical circuit VLSI tools. It is close to impossible with today's tools to compile, manipulate, optimize, and analyze quantum circuits consisting of even a few million gates~\cite{nation2025benchmarking}. Practical quantum circuits, such as~\cite{babbush2018encoding, gidney2021factor, campbell2021early}, include hundreds of millions of high-level quantum gates such as Toffoli, which result in even larger circuits when compiled into Clifford+T.

The compilation and manipulation (e.g. optimization, equivalence checking) of such circuits is being performed in a hierarchic manner, where higher level constructs, e.g. adders, multipliers and QRAM, are used as building blocks. The latter are already optimized for certain costs, such as gate count and depth. Nevertheless, right before its execution on real hardware (at the latest), there exists a point when the entire quantum circuit has to be compiled to native, lower-level gates, and the resulting circuit will have sizes which are impractical for today's tools. Moreover, precise quantum circuit analysis during quantum resource estimation requires the availability of the full circuit.

To the best of our knowledge, there exists no highly parallel, scalable tool which can handle very large quantum circuits. Some of the most scalable quantum circuit design tools can manipulate circuits of thousands of gates (e.g.,~\cite{Sivarajah_2020}), can aggressively optimize very specific circuit types with hundreds of gates (e.g.~\cite{nam2018automated}), or require heuristics for partitioning and rebuilding the circuits (e.g.~\cite{wu2020qgo}).

One of the culprits behind the lack of scalability in today's quantum software is the duration of the standard circuit manipulation operations, such as gate insertion, deletion, replacement, etc. So far, quantum software has relied on high-performance programming languages, such as Rust~\cite{Sivarajah_2020} or Fortran~\cite{nam2018automated} to achieve speed. Fundamentally, the approach taken by most quantum software tools is to store circuits in efficient data structures, such as lists or trees, where the nodes represent the quantum gates of the circuits. 

Depending on the task at hand, these data structures can be more or less advantageous when considering access speed and operation latency. Sequential optimization passes typically benefit from lists for efficient linear traversal, while reinforcement learning-based quantum circuit optimizers~\cite{fösel2021quantum, li2023quarl, moflic2023cost} could exploit tree structures for faster random access to the gates of the circuit during training. Although processing speeds can be further improved through the use of multi-threading, it is generally challenging to write tools that ensure the correctness of the underlying data structures in parallel/distributed scenarios.

Our main contribution is \emph{Pandora}, a software engine built for storing (i.e. caching) and rewriting very large circuits in place. Internally, Pandora represents circuits as doubly linked lists (DLL), which it can process and optimize through custom rewrite rules performed via pattern matching on the stored DLL. Such operations can be applied both sequentially (i.e., linearly iterating over the DLL) and randomly (i.e., accessing gates in random locations of the DLL).

Pandora is built on top of state-of-the-art relational database systems, which allows us to store quantum circuits in database tables for long-term caching. The tables (list data structures) and their associated indices (implemented as trees, hash-maps, etc.) allow us to achieve consistently very high sequential- and random-access speeds for storing, compiling, and manipulating quantum circuits. Moreover, we also get out-of-the-box multi-threading support for correct circuit manipulation operations. We made Pandora HPC-ready by supporting the map-reduce programming model (Section~\ref{sec:hpc}).

Pandora's circuit rewrite engine allows searching for arbitrary gate patterns in the circuit and this enables high-speed compilations (Section~\ref{sec:fermi}) and optimizations (Section~\ref{sec:utility}), low-latency streaming of quantum circuit partitions in resource estimation and quantum HPC~\cite{moflic2025quantumcircuitcachescompressors} scenarios (Section~\ref{sec:part}), as well as a solid foundation for developing scalable and consistent circuit equivalence checkers (Section~\ref{sec:mqt}).

\section{Methods}

We implemented Pandora on top of PostgreSQL v.14~\cite{Postgres} and the rewrite engine is written in the PL/pgSQL dialect, and we use the default transaction isolation level (read committed). We use the Python APIs for interfacing with Qualtran, pyLQTR, Qiskit and TKET.  In the following, we describe the architecture, the core functionalities, and the implementation of Pandora. More technical details are in the Appendix.

\subsection{Architecture}
\label{sec:hpc}

\begin{figure}[!t]
    \centering
    \includegraphics[width=0.75\columnwidth]{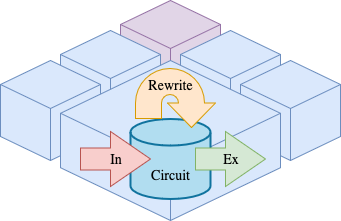}
    \caption{Pandora's architecture. Each box is an independent container in the map-reduce computational model. Blue boxes run Pandora and the magenta boxes control the execution of the blue ones. Each container is executed on a multi-core/-threaded node of an HPC system, and there are independent circuit rewrite threads and insert/extraction threads.}
    \label{fig:framework}
\end{figure}

The architecture of Pandora (Fig.~\ref{fig:framework}) is built around containers, each running an RDBMS instance that stores multiple circuits, using one table for each circuit. This functionality enables the caching of the subcircuits/partitions (Section~\ref{sec:part}). The rewrite engine is loaded in the form of stored procedures within the RDBMS and is accessed through a Python API. 

Containers can be easily configured in a way that enables a map-reduce distributed model of computation: extremely large quantum circuits can be compiled and manipulated by splitting tasks across multiple nodes of an HPC system, where each node runs a Pandora container. The map-reduce model is extremely useful, for example, for full circuit compilations of Shor's algorithm on more than 2048 bits. 

Each node can have multiple processes/threads so that the multi-threading performance of each Pandora container is sped up (Section~\ref{sec:multi-threading}). In the following, we discuss the single-node, multi-threading architecture.

There are two types of threads: 1) the \emph{rewrite} threads, which execute rewrites (see Alg.~\ref{alg:general}); 2) the \emph{insertion/extraction} threads, which import/export the circuit from Pandora or measure circuit-related costs. The latter can be used, for example, to: a) track the circuit costs in time during a rewrite procedure and generate plots similar to the ones from Fig.~\ref{fig:adder_bench} in the Appendix; b) reimport (sub)circuits into external quantum software like Qiskit, Cirq or Quirk.

During rewrites, the rewrite threads in Fig.~\ref{fig:framework} work independently. RDBMS have built-in support for synchronization mechanisms (i.e. locking for concurrent updates) and include multiple ways to enable parallel access and transactions (i.e. atomic operations) on the stored quantum circuit. Transaction synchronization is performed according to the isolation level chosen by the user and, in general, is useful when running concurrent rewrites: a thread must acquire the appropriate locks to ensure that no other thread is concurrently modifying the same rows that correspond to a specific gate template. 

Rewrites within Pandora have access to the entire circuit, yet insertions and extractions into/from Pandora are performed in a streamlined fashion in order to keep the memory requirements low. At insertion/extraction, we collect a fixed-size batch of gates and insert/extract them as a single entity. After insertion/extraction, the process frees the memory associated with the already processed batch and prepares a new batch.

\subsection{Multi-Threaded Rewriting and Performance Model}
\label{sec:model}

Understanding the performance of the Pandora multi-threaded rewrite engine is critical for determining the scalability bottlenecks and the needed future work. The general approach to rewriting circuits is described in Alg.~\ref{alg:general} and we develop the performance model by analyzing the algorithm's steps.

\begin{algorithm}[!t]

\SetKwInOut{Input}{Input}
\SetKwInOut{Output}{Output}

\Input{
  $\textit{pid}$: unique ID of this worker process \\
  $\textit{nproc}$: total number of worker processes \\
  $\textit{pass\_count}$: nr. of passes over the circuit \\
  $\textit{pattern}$: gate pattern to match\\
  $\textit{repl}$: replacement gate sequence}
\Output{
  Rewritten circuit $C$
}

\While{$\textit{pass\_count} > 0$}{
    \ForEach{$\textit{gate}$ in \textit{C} \textbf{where}
      $\textit{gate.id} \bmod \textit{nproc} = \textit{pid}$ 
      \textbf{and} \textit{matches}($\textit{pattern}$)}{
      
      Identify all \textit{neighbor\_gates} required for the pattern\;
      
      Attempt to \textbf{lock} all \textit{neighbor\_gates} and \textit{gate};
      
      \If{\textit{lock acquisition fails for any gate}}{
        \textbf{continue} to next candidate\;
      }

      \textbf{Insert} gates to match pattern if applicable\;
      
      Compute any new link identifiers for the replacement pattern\;
      
      Apply \textbf{update} operations to reconnect neighbors\;
      
      Replace original pattern with replacement gates $repl$\;
      
      \textbf{Delete} any obsolete gates removed by the rewrite\;
         
    \textbf{Commit} transaction to release locks\;
  }
  
  $\textit{pass\_count} \gets \textit{pass\_count} - 1$\;
}

\label{alg:general}
\caption{Multi-threaded rewriting of a circuit assuming a computer with $nproc$ processors. Each process has a unique identifier $pid$ and is running a sequence of $pass\_count$ rewrite passes.}

\end{algorithm}

When rewriting any circuit, Pandora is first \emph{searching} for the patterns that fulfill a certain criterion and can be rewritten, then it is \emph{rewriting} the circuit at the location of the found patterns. Additionally, in a multi-threaded context where multiple threads are rewriting in parallel, there is a need for thread synchronization which is achieved through \emph{communication}.

This implies that the total execution time of a circuit rewrite pass is the sum of three components: the search time $S$, the communication time $C$, and the rewrite time $R$. In the context of rewriting quantum circuits, $S$ is the total time it takes to find all patterns of a specific rewrite in the circuit (\textbf{foreach} loop in Alg.~\ref{alg:general}), $C$ is the total synchronization time of the rewrite threads (\textbf{lock} acquisition mechanisms in Alg.~\ref{alg:general}), and $R$ is the time it takes to rewrite the found templates (i.e. \textbf{update}, \textbf{delete} and \textbf{insert} operations).

The $S, R$ and $C$ times are influenced by the number of $N$ gates in the circuit, the number of $n$ parallel threads, and the probability $p_r$ of encountering a specific template pattern. In general, we assume that $p_r$ is low because. For example, in the case of T-count cancellations in Clifford+T circuits, there will be many opportunities to commute T gates through the circuit, but extremely few possibilities for canceling/removing T gates.

Circuit rewriting is a local operation, such that we can introduce the additional parameters: $s$ is the time to check that a single gate is part of a pattern that can be rewritten, $r$ is the time it takes to rewrite a valid pattern, and $c$ is the synchronization time per rewrite. A general rewrite, such as Alg.~\ref{alg:general}, is traversing the entire circuit of $N$ gates to determine all the valid candidate patterns, but the rewrite is performed only on the $p_r N$ elements which fulfill the rewrite criteria:
\begin{equation*}
    T_1 = S + R = s \times N + r \times (p_rN)
\end{equation*}

The above equation corresponds to the single-threaded case, where the communication cost is zero. However, for $n$ threads, the rewrite time of an entire circuit becomes:
\begin{equation*}
    T_n = S + R + C = s \times N + r \times \frac{p_rN}{n} + c \times (p_rN)
\end{equation*}

Performance bottlenecks are identified by varying $n$ and $p_r$, and this helps with computing the values of $s$, $c$ or $r$ for specific hardware platforms. The speedup obtained by rewriting the circuit with $n$ threads is:
\begin{equation*}
    \frac{T_1}{T_n} = \frac{s N + r (p_rN)}{ s N + r \frac{p_rN}{n} + c (p_rN)} \\
    = \frac{s + rp_r}{s + r\frac{p_r}{n} + cp_r}
\end{equation*}

We can relate our speed-up to a form similar to Amdahl's law, and this shows that we are limited by the sequential component (search $S$ and communication $C$) when rewriting circuits with parallel threads.

This performance model can be used to explain the behavior of Pandora. In the Results section, we show in Fig.~\ref{fig:utility} that there is an advantage in using Pandora (even if just sequentially and single-threaded) for low probabilities $p_r$. Having a n RDBMS as an engine allows for a reduced search time and lowers the sequential fraction of the computation. Fig.~\ref{fig:multi} illustrates the multi-threading performance of Pandora appears at least in the following scenarios: (a) almost ideal speed-ups are obtained with parallel circuit transpilation: $s$ and $c$ are nonexistent here; (b) multi-threading for applying template from Fig.~\ref{fig:rules}g in $10\%$ of the total gate count shows that good speed-ups can still be achieved even for large values of $c$.

\subsection{Circuit Rewrite Methods}
\label{sec:rewrites}

Quantum circuit rewriting involves applying template-based circuit rewrite rules~\cite{maslov2005quantum} in a manner that reduces a cost of interest (e.g., gate count, circuit depth, connectivity gate overhead, or gate cost in hardware). Template rewrite rules (e.g. \ref{fig:rules}) are a set of known circuit equivalences that, if combined in an efficient manner, can optimize quantum circuits by allowing either gate cancellation or gate parallelization to occur.

Various heuristic-based methods have been proposed for effective quantum circuit rewriting. Most of these heuristics rely on applying local or global template rewrite rules in order to meet a specific optimization criterion. A recent popular choice is Reinforcement Learning (RL)~\cite{fösel2021quantum, li2023quarl, moflic2023cost}, in which optimization criteria are encoded in a reward function that the RL agent is expected to minimize by exploring different strategies.

However, for large-scale circuits, memory can quickly become a bottleneck. Inefficient access patterns or swapping lead to performance degradation and higher optimization costs~\cite{paler2022energy}. Each rewrite rule within Pandora is implemented as a separate, atomic transaction and operates on database tables (Section~\ref{sec:database}, Appendix).

We define the operation $\sigma_{\text{condition}}(\mathcal{C})$, which selects rows from the table $\mathcal{C}$ that satisfy a given predicate. The output is a new table $\mathcal{C'}$, that consists only of the selected rows. For any row $r$, we use $r[\textit{prop}]$ to refer to the value of the column \textit{prop}, and $r[\textit{prop}] \leftarrow \textit{nval}$ to denote an assignment operation. Relational algebra does not include random sampling,  therefore we extend the notation slightly to describe the semantics of our rewrite rules:

\begin{itemize}
    \item \textbf{Select} a gate with identifier $p_1$: $\sigma_{\mathcal{C}}(\textit{id} = p_1)$
    
    \item \textbf{Randomly select} a gate of type $t_1$: $\sigma_{\mathcal{C}}(\text{rand}, \, \textit{gate\_type} = t_1)$
    
    \item \textbf{Insert} a new gate $E$: $\mathcal{C'} \leftarrow \mathcal{C} \cup E$
    
    \item \textbf{Delete} a gate by ID: $\mathcal{C'} \leftarrow \mathcal{C} \setminus \sigma_{\mathcal{C}}(\textit{id} = p_1)$
\end{itemize}

Each rewrite rule is executed in a transaction composed of a sequence of operations such as those described above. The rewrites are implemented as stored procedures -- a collection of database (SQL) queries written as a single database transaction. Stored procedures have several advantages -- are compiled only once and behave like parameterized, executable-form function calls. Stored procedures improve performance, as a single call is sent to the database server instead of multiple SQL queries. Moreover, stored procedures can run in parallel and modify the table entries concurrently. Deadlock avoidance and correctness during the rewriting process is enabled by carefully selecting the candidate rows that match a template.

For compilation (e.g., transpiling Toffoli circuits into Clifford+T) and optimization purposes, the threads in Fig.~\ref{fig:framework} execute a single or a sequence of rewrites until a certain criterion is met. Alg.~\ref{alg:general} is the pseudocode executed within each thread. Rewrite rules do not depend on a gate's position in the circuit. Therefore, in a worst-case, the application of each rewrite parameterized by a particular gate id or type will require searching the entire table. Select operations will be sped up by using fast indexing of tables.

\subsection{Partitioning}
\label{sec:part}

Quantum circuits such as the ones from quantum chemistry (e.g., Fermi-Hubbard circuits) or Shor's algorithm circuits exceed the capabilities of today's practical settings. Recently, several hardware roadmaps~\cite{saadatmand2024fault} have proposed distributed or aggregated quantum architectures. In such architectures, multiple quantum processing units (QPU) -- which have a limited qubit count due to technical or engineering constraints -- are interconnected in order to support the execution of very large-scale computations.

QPUs do not necessarily have homogeneous internal architectures -- some QPUs are envisioned to be used for distilling and distributing T states, while other QPUs are envisioned for the Clifford part of the computation. To this end, compiling quantum circuits for distributed architectures requires partitioning the computation into subcircuits which are aware of the type and the resource constraints of each QPU in the system.

A \emph{partition} is a subcircuit compatible with the type and constraints of a particular QPU type. \emph{Partitioning} is the process of automatically extracting such subcircuits from large-scale quantum circuits. 

Nevertheless, partitions are difficult to optimize with respect to metrics such as $T$-count and $T$-depth. Determining and processing a partition is a time-consuming task and involves transpiling the partition to the native gate set of the quantum computer~\cite{baker2020time, brandhofer2023optimal} and further optimizing it to the constraints of the quantum compute unit. There is already a rich literature on circuit cutting and partitioning, yet none of those works focus on the speed of partitioning or the pure speeds of communicating the partitions towards the computational units.

We implemented a partitioning algorithm on top of Pandora. Our partitioning method is based on a union-find routing on the quantum circuit represented as a DLL. The quantum circuit is imported/cached in Pandora, and the partition search is taking place in memory on the DAGs extracted from Pandora. For efficiency, partitions are processed in fixed-sized batches of the original DAG. For example, we can choose a batch size of $10^6$ gates, and then extract and partition the the batch into components with a given depth $d$ and a maximum of $t$ gates at a time. Once a batch is fully partitioned, the algorithm proceeds with the next batch. 

The complexity of our algorithm is linear in the number of DAG vertices. The union-find algorithm has a space complexity of $\mathcal{O}(n)$, where $n$ is the number of nodes in the DAG and a time complexity of $\mathcal{O}(\alpha(n))$, which is practically constant even for very large inputs -- $\alpha$ grows extremely slowly.

\section{Results}
\label{sec:results}

We have designed, implemented, open-sourced~\footnote{\url{https://github.com/ioanamoflic/pandora}} and benchmarked Pandora. The benchmark circuits are generated by state-of-the-art quantum software, such as Cirq 1.4, Qiskit 2.1, Qualtran 0.6, pyLIQTR 1.3.3 and pytket 2.8. For visual inspection of the small circuits, we can easily export and visualize circuits with Quirk. Our benchmarks were executed on a Theadripper 7960X 24-core machine with 256GB RAM. The HPC capabilities of Pandora were tested on the LUMI supercomputer.

\begin{figure}[!t]
    \centering
    \includegraphics[width=0.99\columnwidth]{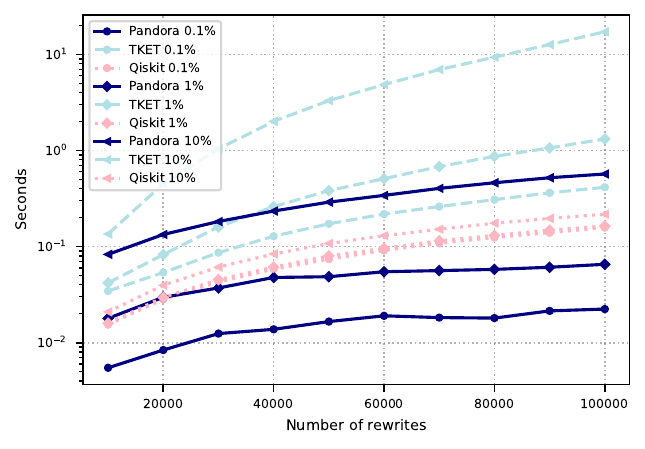}
    \caption{Pandora outperforms both TKET and Qiskit on circuits of more than $10^4$ CNOT gates where a rewrite is applied less than 10\% of the time (i.e. Fig.~\ref{fig:rules} g) in Appendix). The advantage of Pandora stems from the fast pattern search times. For example, for medium-sized circuits, both Qiskit ans TKET spend most of the time searching for the pattern instead of rewriting.}
    \label{fig:utility}
\end{figure}

\emph{Synthetic quantum circuits} are used for stressing Pandora's rewrite engine, while at the same time not requiring the implementation of complex compilation/optimization algorithms that depend on the structure of practical quantum circuits. For example, to stress test the speed of canceling quantum gates in quantum addition circuits, one would need to implement complex methods such as those of~\cite{nam2018automated}. Nevertheless, Pandora's pure speed is applicable for practical circuits, and we show that we can optimize adders by rewriting and canceling gates at random positions in the circuits.

\emph{Practical quantum circuits} are used to measure caching and circuit partitioning performance for the purpose of quantum resource estimation and distributed QHPCs. Additionally, Pandora's storage and caching performance is used to compile extremely large circuits such as Shor's algorithm.

\subsection{Utility Threshold}
\label{sec:utility}

\begin{figure}[t!]
\includegraphics[width=0.4\textwidth]{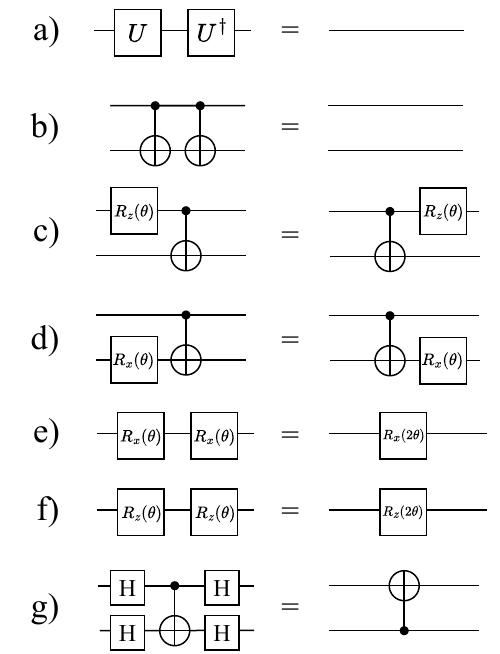}
\centering
\caption{Circuit rewrites for single qubit rotations and CNOTs. From top to bottom: a) a unitary followed by its conjugate transposed is equivalent to identity. This rule will also be referred to in this manuscript as the HADAMARD rule for $U=H$; b) two consecutive CNOT gates cancel out; c) Z-rotations commute with the control of CNOTs; d) X-rotations commute with the target of the CNOT; e) two consecutive $R_x(\theta)$ rotations yield a single $R_x$ rotation with the angle doubled; f) two consecutive $R_z(\theta)$ rotations yield a single $R_z$ rotation with the angle doubled; g) the direction of a CNOT gate can be reversed if it is "surrounded" by Hadamard gates on every qubit. This rule will also be referred to in this manuscript as the REVERSE rule.}
\label{fig:rules}
\end{figure}

Pandora's rewriting performance is compared to TKET~\cite{Sivarajah_2020} and IBM Qiskit, which are two of the fastest open-source quantum circuit quantum software currently available. We generate random circuits with $N$ (in range $10^4$ to $10^5$) CNOTS and a fixed qubit size. We select a probability $p_r$ ($0.1\%$, $1\%$, and $10\%$) and flip $p_r \times N$ CNOT gates using the reverse of rewrite Fig.~\ref{fig:rules} g). For both TKET and Qiskit, the goal is to traverse the circuit sequentially and to rewrite each flipped CNOT using the rewrite rule (Fig.~\ref{fig:rules} g). In Pandora, we use a sequential variant of Alg.~\ref{alg:general} and apply the same procedure. 

The results are summarized in Fig.~\ref{fig:utility}. Pandora becomes especially useful when the patterns we search for are rarely encountered in the circuit -- and this is the case almost all the time for quantum circuit optimisatin. For example, for $p_r = 0.1$ and $p_r = 1$, Pandora outperforms both Qiskit and TKET.

\subsection{Multi-Threading Performance}
\label{sec:multi-threading}

Pandora is multi-threaded, and we report the speedups obtained by running the TKET/Qiskit benchmarks and Fermi-Hubbard $50 \times 50$ compilations on $2,4, \ldots, 16$ cores. The Fermi-Hubbard $50 \times 50$ circuit is generated by \texttt{pyLIQTR} and the resulting $514E+6$ Clifford+T gates are stored into Pandora at speeds of up to $270E+3$ gates/second when using 16 cores. Additional multi-threading results are reported in Sections~\ref{sec:fermi}, and~\ref{sec:mqt}.

\begin{figure}[!t]
    \centering
    \includegraphics[width=0.9\columnwidth]{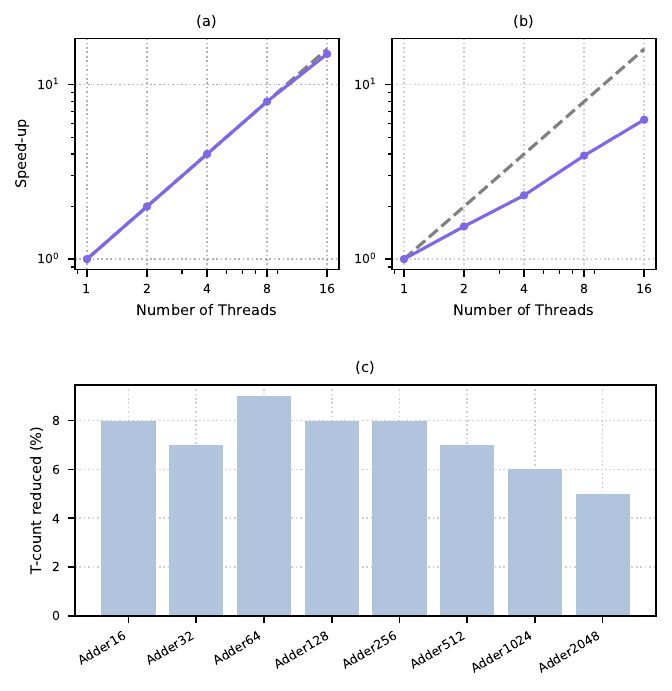}
    
    \caption{Speed-up achieved with multi-threading are close to ideal. The dotted gray line shows the ideal speed-up, and the blue one shows the speed-up obtained by the multi-threaded rewrites (100 000 CNOTs circuit, with $p_r = 10\%$ ) and multi-threaded compilations (Fermi-Hubbard 50x50). The results of multi-threaded rewriting of adder circuits of 8-2048 bits. The threads were assigned one of the rewrite from Fig.~\ref{fig:rules}.}
    \label{fig:multi}
\end{figure}

The previous benchmarks use the same rewrite rule across the threads. However, it is possible to have multiple threads with each of them executing a different type of rewrite. Such an approach would be similar to performing some kind of quantum circuit optimization.

Without implementing any optimization heuristic, and being interested just in the pure speed of our rewrite engine, we use quantum addition circuits from~\cite{nam2018automated} and let 48 threads implement all the rewrites from Fig.~\ref{fig:rules}. We run the threads for 600 seconds and record the T-count both before and after the procedure. The results can be visualized in Fig.~\ref{fig:multi}. The time evolution of different gate counts for all the adders are presented in Fig.~\ref{fig:adder_bench} of the Appendix.

\subsection{Caches for Real-time Compilation}
\label{sec:fermi}

Caches improve the performance and latency of an application by storing frequently accessed data in fast memory storage. Pandora supports the long-term storage of precompiled sub-circuits (e.g. partitions), such that it can operate as a cache during the compilation of ultra-large-scale circuits. We report the performance of inserting and caching circuits in Pandora and then reusing the stored circuits as needed during the complete compilation of Shor's algorithm circuits and the partitioning of Fermi-Hubbard.

\begin{figure}[!t]
    \centering
    \includegraphics[width=0.9\columnwidth]{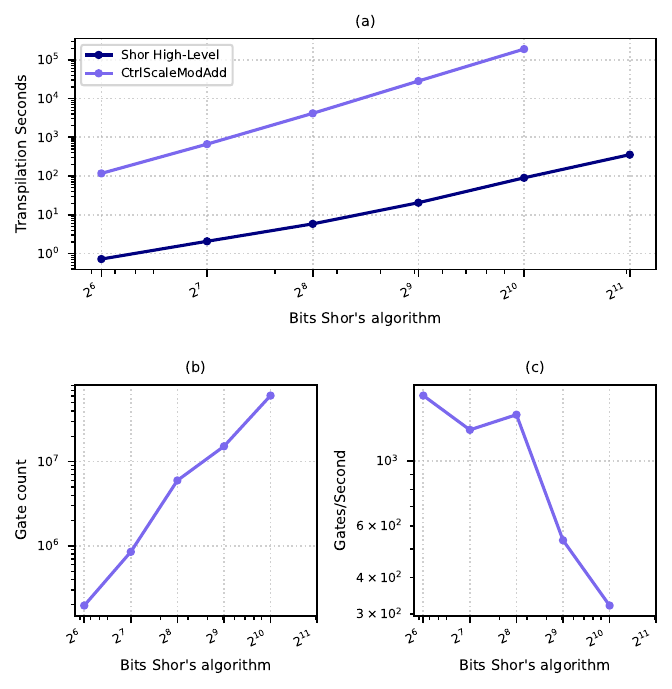}
    \caption{Speed of decomposing the \texttt{CtrlScaleModAdd} Bloq with various input sizes. (a) The transpilation time in seconds for the high-level Bloq and its full transpilation time to a reduced gate set. (b) Final gate count of \texttt{CtrlScaleModAdd} when decomposed down to at most three-qubit gates. (c) The number of gates processed per second decreases with input size -- this is due to the computational overheads from the source quantum software used for decomposing the circuits.}
    \label{fig:shor}
\end{figure}

We fully compile Shor's algorithm circuits using Qualtran (Fig.~\ref{fig:shor}). Because the latter is still under development it does not include a lower-level decomposition of \texttt{CtrlScaleModAdd} and we had to use one from \texttt{pyLIQTR}. For benchmarking purposes, we compile the complete high-level Shor's circuit, and decompose and cache into Pandora a single \texttt{CtrlScaleModAdd} into CNOT, Toffoli, CSWAP and single-qubit Clifford gates.

One of the largest utility-scale circuits decomposed with Pandora was a $100\times 100$ Fermi-Hubbard instance obtained from \texttt{pyLIQTR} which had on the order of $7.7E+9$ Clifford+T gates.

\subsection{High-Throughput Circuit Partitioning}
\label{sec:part}

Partitions exist naturally in most of the quantum circuits (e.g. Shor's algorithm, Fermi-Hubbard simulations) due to their modular nature. For example, Shor's algorithm circuit is built out of multipliers which are built out of adders. Therefore, the hierarchic structure of the high-level circuits can be used for optimization purposes: a more efficient adder will automatically result in a more efficient implementation of Shor's algorithm. Nevertheless, a circuit's modules might still be too large abstractions for compiling and executing computations on heterogeneous QHPCs. This is due to the fact that QHPC processing units will have limited computational resources (e.g. number of qubits, gate-count and gate-depth, T-gates). As a result, circuit/module partitions need to be computed and streamed as fast as possible towards the QHPC units.

\begin{figure}[!t]
    \centering
    \includegraphics[width=0.99\columnwidth]{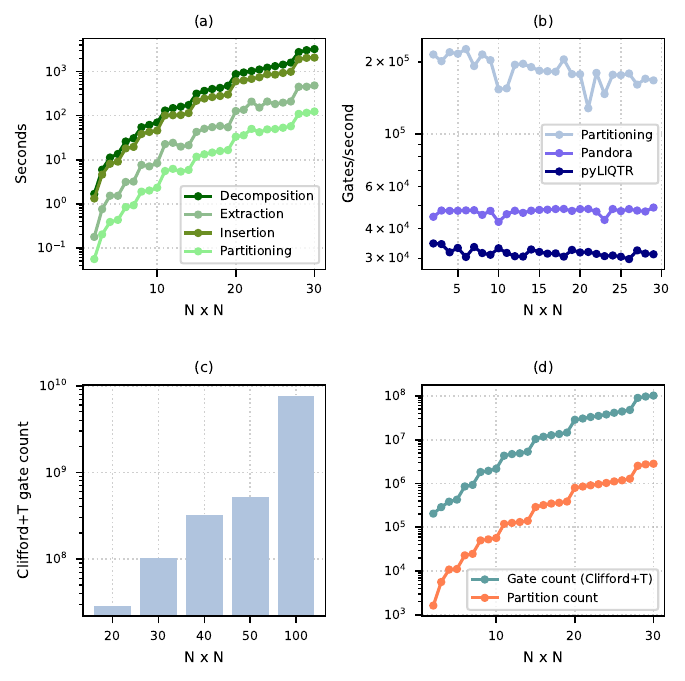}
    \caption{Compilation for Fermi-Hubbard $N \times N$ circuits. (a) Durations for the Pandora pipeline in order of execution: decomposition time in \texttt{pyLIQTR}, insertion, partitioning and extraction times; (b) The number of gates processed per second at different stages of the pipeline - once a circuit is decomposed and stored, the partitioning algorithm achieves very high speeds, about five times faster than \texttt{pyLIQTR} decomposition; (c) Total Clifford+T gate counts after compilation. (d) Number of partitions when setting the constraints to $200$ gates and at most $100$ T-gates.}
    \label{fig:partition}
\end{figure}

We implemented a fast circuit partition based on \emph{union-find} (Section~\ref{sec:part}) that can automatically build partitions according to specific constraints of the compute units, such as T-count, gate-count, gate-depth, number of I/O, etc. In order to partition, we build a pipeline of operations: we compile and decompose to Clifford+T the circuits with pyLIQTR, we insert, store and cache the resulting circuits into Pandora, after which we partition the stored circuits and extract (stream) the partitions out of Pandora. Extraction is the process of assembling the Pandora information into a standardized format such as OpenQASM. The throughput of the extraction, in terms of gates/second, is a function of the partitioning speed -- slow partitioning will bottleneck the extraction speed. 

We show that high-throughput partitioning is possible. The obtained compilation and partitioning speeds are shown Fig.\ref{fig:partition}a: the transpilation from \texttt{pyLIQTR} and the insertion are the most time-consuming pipeline stages. Partitioning and extraction are the most efficient stages. 

The sensitivity of our partitioning method to the gate-depth and T-count constraints is illustrated in Fig.~\ref{fig:partition}d. For fixed depth and T count, the algorithm behaves as expected: the partition processing speed increases with larger circuit instances.

\subsection{Equivalence Checking of Quantum Circuits}
\label{sec:mqt}

The functional equivalence checking of quantum circuits is the problem of determining whether two quantum circuits implement the same computation~\cite{lewis2023formal, yamashita2010fast, burgholzer2020advanced}. For particular classes of quantum circuits this can be solved efficiently~\cite{van2020zx, elliott2008graphical, paler2018specification} but, in general, determining if two quantum circuits are functionally equivalent is a highly complex task. This type of checking becomes necessary to ensure that an optimized circuit is implementing the same computation as the original one. However, optimization might introduce faults into the circuit, such as missing gates or qubits. The latter are easier to catch, while the first more difficult. 

The general approach to determine equivalence between two circuits $C_1$ (the original circuit)  and $C_2$ (the optimized circuit) is to show that $C_2C_1^\dagger = I$ or $C_1C_2^\dagger = I$. Whenever this is not the case, the circuits are not equivalent and $C_2 \neq C_1$. Various approaches have been proposed for equivalence checking, and using rewrite engines is one of the original ones~\cite{yamashita2010fast}.

\begin{figure}[!t]
    \centering
    \includegraphics[width=0.99\columnwidth]{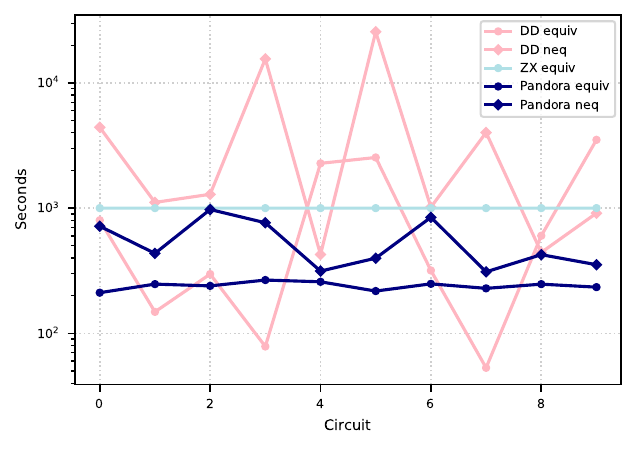}
    \caption{Rewrite speed of Pandora (dark blue) for the purpose of implementing a equivalence checking. We generated 10 random circuits (index on the horizontal axis) of 32 qubits and 27000 gates and measured the time it takes the decision-diagram MQT.QCEC tool (pink) to show (non-)equivalence. All checkers were run with a timeout of 1000 seconds. The ZX checker (light blue) timed out always before deciding. The Pandora non-equivalence checking method is probabilistic, but for these benchmark circuits it never timed out. The Pandora equivalence checking routine did never time out.}
    \label{fig:mqt}
\end{figure}

In this paper, we are concerned with the speed of determining equivalence without developing a general, complete method for equivalence checking. This is to show that there is potential in using Pandora for developing multi-threaded, rewrite-based equivalence checkers. To this end, we restrict our numeric analysis to initial circuits $C_1$ of $q$ qubits having $q^3$  (CNOT) gates applied randomly. We are testing two cases: a) when $C_1 = C_2$, we generate $C_2$ simply as a copy of $C_1$; b) when $C_1 \neq C_2$ we generate $C_2$ by removing a random gate from $C_1$. 

The construction of our benchmark circuits is the same to the one in~\cite{burgholzer2020advanced}, and we benchmark against the same state-of-the-art tools, MQT QCEC. Our results (Fig.~\ref{fig:mqt}) show that for $q > 30$, Pandora can perform all the pairwise cancellations from $C_1C_2^\dagger$ faster than the general equivalence checker of~\cite{burgholzer2020advanced}.

\section{Conclusion}
\label{sec:concl}

Pandora demonstrates a significant advance in the field of quantum circuit manipulation by leveraging a novel architecture based on state-of-the-art relational database management systems. Our work addresses the critical scalability gap in quantum software, enabling the compilation and analysis of circuits far exceeding the capabilities of existing tools. The system's capacity to handle circuits with billions of gates, coupled with its multi-threaded, high-performance-computing-enabled design, allows for efficient processing of real-world problems such as 1024-bit Shor's algorithm and 100×100 Fermi-Hubbard instances.

Performance analysis shows a clear advantage over contemporary frameworks like TKET and Qiskit for circuits with more than 10,000 gates and for equivalence checking on circuits with more than 32 qubits. This is particularly evident in scenarios where rewrite patterns are rare. Pandora's streamed decomposition pipeline, which processes circuits in batches, effectively mitigates the memory footprint bottleneck, a common challenge for large circuits. The implementation on PostgreSQL and its map-reduce capabilities for HPC systems further underscore its potential for deployment in distributed quantum architectures.

\begin{acknowledgments}
The authors thank Laurenz Albe for his extremely valuable Stackoverflow answers on improving PostgreSQL performance. This research was developed in part with funding from the Defense Advanced Research Projects Agency [under the Quantum Benchmarking (QB) program under award no. HR00112230006 and HR001121S0026 contracts]. The views, opinions and/or findings expressed are those of the author(s) and should not be interpreted as representing the official views or policies of the Department of Defense or the U.S. Government.
\end{acknowledgments}

\bibliography{__main}

\begin{thebibliography}{31}%
\makeatletter
\providecommand \@ifxundefined [1]{%
 \@ifx{#1\undefined}
}%
\providecommand \@ifnum [1]{%
 \ifnum #1\expandafter \@firstoftwo
 \else \expandafter \@secondoftwo
 \fi
}%
\providecommand \@ifx [1]{%
 \ifx #1\expandafter \@firstoftwo
 \else \expandafter \@secondoftwo
 \fi
}%
\providecommand \natexlab [1]{#1}%
\providecommand \enquote  [1]{``#1''}%
\providecommand \bibnamefont  [1]{#1}%
\providecommand \bibfnamefont [1]{#1}%
\providecommand \citenamefont [1]{#1}%
\providecommand \href@noop [0]{\@secondoftwo}%
\providecommand \href [0]{\begingroup \@sanitize@url \@href}%
\providecommand \@href[1]{\@@startlink{#1}\@@href}%
\providecommand \@@href[1]{\endgroup#1\@@endlink}%
\providecommand \@sanitize@url [0]{\catcode `\\12\catcode `\$12\catcode `\&12\catcode `\#12\catcode `\^12\catcode `\_12\catcode `\%12\relax}%
\providecommand \@@startlink[1]{}%
\providecommand \@@endlink[0]{}%
\providecommand \url  [0]{\begingroup\@sanitize@url \@url }%
\providecommand \@url [1]{\endgroup\@href {#1}{\urlprefix }}%
\providecommand \urlprefix  [0]{URL }%
\providecommand \Eprint [0]{\href }%
\providecommand \doibase [0]{https://doi.org/}%
\providecommand \selectlanguage [0]{\@gobble}%
\providecommand \bibinfo  [0]{\@secondoftwo}%
\providecommand \bibfield  [0]{\@secondoftwo}%
\providecommand \translation [1]{[#1]}%
\providecommand \BibitemOpen [0]{}%
\providecommand \bibitemStop [0]{}%
\providecommand \bibitemNoStop [0]{.\EOS\space}%
\providecommand \EOS [0]{\spacefactor3000\relax}%
\providecommand \BibitemShut  [1]{\csname bibitem#1\endcsname}%
\let\auto@bib@innerbib\@empty
\bibitem [{\citenamefont {Nation}\ \emph {et~al.}(2025)\citenamefont {Nation}, \citenamefont {Saki}, \citenamefont {Brandhofer}, \citenamefont {Bello}, \citenamefont {Garion}, \citenamefont {Treinish},\ and\ \citenamefont {Javadi-Abhari}}]{nation2025benchmarking}%
  \BibitemOpen
  \bibfield  {author} {\bibinfo {author} {\bibfnamefont {P.~D.}\ \bibnamefont {Nation}}, \bibinfo {author} {\bibfnamefont {A.~A.}\ \bibnamefont {Saki}}, \bibinfo {author} {\bibfnamefont {S.}~\bibnamefont {Brandhofer}}, \bibinfo {author} {\bibfnamefont {L.}~\bibnamefont {Bello}}, \bibinfo {author} {\bibfnamefont {S.}~\bibnamefont {Garion}}, \bibinfo {author} {\bibfnamefont {M.}~\bibnamefont {Treinish}},\ and\ \bibinfo {author} {\bibfnamefont {A.}~\bibnamefont {Javadi-Abhari}},\ }\bibfield  {title} {\bibinfo {title} {Benchmarking the performance of quantum computing software for quantum circuit creation, manipulation and compilation},\ }\href@noop {} {\bibfield  {journal} {\bibinfo  {journal} {Nature Computational Science}\ ,\ \bibinfo {pages} {1}} (\bibinfo {year} {2025})}\BibitemShut {NoStop}%
\bibitem [{\citenamefont {Babbush}\ \emph {et~al.}(2018)\citenamefont {Babbush}, \citenamefont {Gidney}, \citenamefont {Berry}, \citenamefont {Wiebe}, \citenamefont {McClean}, \citenamefont {Paler}, \citenamefont {Fowler},\ and\ \citenamefont {Neven}}]{babbush2018encoding}%
  \BibitemOpen
  \bibfield  {author} {\bibinfo {author} {\bibfnamefont {R.}~\bibnamefont {Babbush}}, \bibinfo {author} {\bibfnamefont {C.}~\bibnamefont {Gidney}}, \bibinfo {author} {\bibfnamefont {D.~W.}\ \bibnamefont {Berry}}, \bibinfo {author} {\bibfnamefont {N.}~\bibnamefont {Wiebe}}, \bibinfo {author} {\bibfnamefont {J.}~\bibnamefont {McClean}}, \bibinfo {author} {\bibfnamefont {A.}~\bibnamefont {Paler}}, \bibinfo {author} {\bibfnamefont {A.}~\bibnamefont {Fowler}},\ and\ \bibinfo {author} {\bibfnamefont {H.}~\bibnamefont {Neven}},\ }\bibfield  {title} {\bibinfo {title} {{Encoding electronic spectra in quantum circuits with linear T complexity}},\ }\href@noop {} {\bibfield  {journal} {\bibinfo  {journal} {Physical Review X}\ }\textbf {\bibinfo {volume} {8}},\ \bibinfo {pages} {041015} (\bibinfo {year} {2018})}\BibitemShut {NoStop}%
\bibitem [{\citenamefont {Gidney}\ and\ \citenamefont {Eker{\aa}}(2021)}]{gidney2021factor}%
  \BibitemOpen
  \bibfield  {author} {\bibinfo {author} {\bibfnamefont {C.}~\bibnamefont {Gidney}}\ and\ \bibinfo {author} {\bibfnamefont {M.}~\bibnamefont {Eker{\aa}}},\ }\bibfield  {title} {\bibinfo {title} {{How to factor 2048 bit RSA integers in 8 hours using 20 million noisy qubits}},\ }\href@noop {} {\bibfield  {journal} {\bibinfo  {journal} {Quantum}\ }\textbf {\bibinfo {volume} {5}},\ \bibinfo {pages} {433} (\bibinfo {year} {2021})}\BibitemShut {NoStop}%
\bibitem [{\citenamefont {Campbell}(2021)}]{campbell2021early}%
  \BibitemOpen
  \bibfield  {author} {\bibinfo {author} {\bibfnamefont {E.~T.}\ \bibnamefont {Campbell}},\ }\bibfield  {title} {\bibinfo {title} {{Early fault-tolerant simulations of the Hubbard model}},\ }\href@noop {} {\bibfield  {journal} {\bibinfo  {journal} {Quantum Science and Technology}\ }\textbf {\bibinfo {volume} {7}},\ \bibinfo {pages} {015007} (\bibinfo {year} {2021})}\BibitemShut {NoStop}%
\bibitem [{\citenamefont {Sivarajah}\ \emph {et~al.}(2020)\citenamefont {Sivarajah}, \citenamefont {Dilkes}, \citenamefont {Cowtan}, \citenamefont {Simmons}, \citenamefont {Edgington},\ and\ \citenamefont {Duncan}}]{Sivarajah_2020}%
  \BibitemOpen
  \bibfield  {author} {\bibinfo {author} {\bibfnamefont {S.}~\bibnamefont {Sivarajah}}, \bibinfo {author} {\bibfnamefont {S.}~\bibnamefont {Dilkes}}, \bibinfo {author} {\bibfnamefont {A.}~\bibnamefont {Cowtan}}, \bibinfo {author} {\bibfnamefont {W.}~\bibnamefont {Simmons}}, \bibinfo {author} {\bibfnamefont {A.}~\bibnamefont {Edgington}},\ and\ \bibinfo {author} {\bibfnamefont {R.}~\bibnamefont {Duncan}},\ }\bibfield  {title} {\bibinfo {title} {{t|ket⟩: a retargetable compiler for NISQ devices}},\ }\href {https://doi.org/10.1088/2058-9565/ab8e92} {\bibfield  {journal} {\bibinfo  {journal} {Quantum Science and Technology}\ }\textbf {\bibinfo {volume} {6}},\ \bibinfo {pages} {014003} (\bibinfo {year} {2020})}\BibitemShut {NoStop}%
\bibitem [{\citenamefont {Nam}\ \emph {et~al.}(2018)\citenamefont {Nam}, \citenamefont {Ross}, \citenamefont {Su}, \citenamefont {Childs},\ and\ \citenamefont {Maslov}}]{nam2018automated}%
  \BibitemOpen
  \bibfield  {author} {\bibinfo {author} {\bibfnamefont {Y.}~\bibnamefont {Nam}}, \bibinfo {author} {\bibfnamefont {N.~J.}\ \bibnamefont {Ross}}, \bibinfo {author} {\bibfnamefont {Y.}~\bibnamefont {Su}}, \bibinfo {author} {\bibfnamefont {A.~M.}\ \bibnamefont {Childs}},\ and\ \bibinfo {author} {\bibfnamefont {D.}~\bibnamefont {Maslov}},\ }\bibfield  {title} {\bibinfo {title} {{Automated optimization of large quantum circuits with continuous parameters}},\ }\href@noop {} {\bibfield  {journal} {\bibinfo  {journal} {npj Quantum Information}\ }\textbf {\bibinfo {volume} {4}},\ \bibinfo {pages} {23} (\bibinfo {year} {2018})}\BibitemShut {NoStop}%
\bibitem [{\citenamefont {Wu}\ \emph {et~al.}(2020)\citenamefont {Wu}, \citenamefont {Davis}, \citenamefont {Chong},\ and\ \citenamefont {Iancu}}]{wu2020qgo}%
  \BibitemOpen
  \bibfield  {author} {\bibinfo {author} {\bibfnamefont {X.-C.}\ \bibnamefont {Wu}}, \bibinfo {author} {\bibfnamefont {M.~G.}\ \bibnamefont {Davis}}, \bibinfo {author} {\bibfnamefont {F.~T.}\ \bibnamefont {Chong}},\ and\ \bibinfo {author} {\bibfnamefont {C.}~\bibnamefont {Iancu}},\ }\bibfield  {title} {\bibinfo {title} {{QGo: Scalable quantum circuit optimization using automated synthesis}},\ }\href@noop {} {\bibfield  {journal} {\bibinfo  {journal} {arXiv preprint arXiv:2012.09835}\ } (\bibinfo {year} {2020})}\BibitemShut {NoStop}%
\bibitem [{\citenamefont {Fösel}\ \emph {et~al.}(2021)\citenamefont {Fösel}, \citenamefont {Niu}, \citenamefont {Marquardt},\ and\ \citenamefont {Li}}]{fösel2021quantum}%
  \BibitemOpen
  \bibfield  {author} {\bibinfo {author} {\bibfnamefont {T.}~\bibnamefont {Fösel}}, \bibinfo {author} {\bibfnamefont {M.~Y.}\ \bibnamefont {Niu}}, \bibinfo {author} {\bibfnamefont {F.}~\bibnamefont {Marquardt}},\ and\ \bibinfo {author} {\bibfnamefont {L.}~\bibnamefont {Li}},\ }\href@noop {} {\bibinfo {title} {{Quantum circuit optimization with deep reinforcement learning}}} (\bibinfo {year} {2021}),\ \Eprint {https://arxiv.org/abs/2103.07585} {arXiv:2103.07585 [quant-ph]} \BibitemShut {NoStop}%
\bibitem [{\citenamefont {Li}\ \emph {et~al.}(2023)\citenamefont {Li}, \citenamefont {Peng}, \citenamefont {Mei}, \citenamefont {Lin}, \citenamefont {Wu}, \citenamefont {Padon},\ and\ \citenamefont {Jia}}]{li2023quarl}%
  \BibitemOpen
  \bibfield  {author} {\bibinfo {author} {\bibfnamefont {Z.}~\bibnamefont {Li}}, \bibinfo {author} {\bibfnamefont {J.}~\bibnamefont {Peng}}, \bibinfo {author} {\bibfnamefont {Y.}~\bibnamefont {Mei}}, \bibinfo {author} {\bibfnamefont {S.}~\bibnamefont {Lin}}, \bibinfo {author} {\bibfnamefont {Y.}~\bibnamefont {Wu}}, \bibinfo {author} {\bibfnamefont {O.}~\bibnamefont {Padon}},\ and\ \bibinfo {author} {\bibfnamefont {Z.}~\bibnamefont {Jia}},\ }\href@noop {} {\bibinfo {title} {{Quarl: A Learning-Based Quantum Circuit Optimizer}}} (\bibinfo {year} {2023}),\ \Eprint {https://arxiv.org/abs/2307.10120} {arXiv:2307.10120 [quant-ph]} \BibitemShut {NoStop}%
\bibitem [{\citenamefont {Moflic}\ and\ \citenamefont {Paler}(2023)}]{moflic2023cost}%
  \BibitemOpen
  \bibfield  {author} {\bibinfo {author} {\bibfnamefont {I.}~\bibnamefont {Moflic}}\ and\ \bibinfo {author} {\bibfnamefont {A.}~\bibnamefont {Paler}},\ }\bibfield  {title} {\bibinfo {title} {{Cost Explosion for Efficient Reinforcement Learning Optimisation of Quantum Circuits}},\ }in\ \href@noop {} {\emph {\bibinfo {booktitle} {2023 IEEE International Conference on Rebooting Computing (ICRC)}}}\ (\bibinfo {organization} {IEEE},\ \bibinfo {year} {2023})\ pp.\ \bibinfo {pages} {1--5}\BibitemShut {NoStop}%
\bibitem [{\citenamefont {Moflic}\ \emph {et~al.}(2025)\citenamefont {Moflic}, \citenamefont {Robertson}, \citenamefont {Devitt},\ and\ \citenamefont {Paler}}]{moflic2025quantumcircuitcachescompressors}%
  \BibitemOpen
  \bibfield  {author} {\bibinfo {author} {\bibfnamefont {I.}~\bibnamefont {Moflic}}, \bibinfo {author} {\bibfnamefont {A.}~\bibnamefont {Robertson}}, \bibinfo {author} {\bibfnamefont {S.~J.}\ \bibnamefont {Devitt}},\ and\ \bibinfo {author} {\bibfnamefont {A.}~\bibnamefont {Paler}},\ }\href {https://arxiv.org/abs/2507.20677} {\bibinfo {title} {Quantum circuit caches and compressors for low latency, high throughput computing}} (\bibinfo {year} {2025}),\ \Eprint {https://arxiv.org/abs/2507.20677} {arXiv:2507.20677 [quant-ph]} \BibitemShut {NoStop}%
\bibitem [{\citenamefont {Stonebraker}\ \emph {et~al.}(1990)\citenamefont {Stonebraker}, \citenamefont {Rowe},\ and\ \citenamefont {Hirohama}}]{Postgres}%
  \BibitemOpen
  \bibfield  {author} {\bibinfo {author} {\bibfnamefont {M.}~\bibnamefont {Stonebraker}}, \bibinfo {author} {\bibfnamefont {L.}~\bibnamefont {Rowe}},\ and\ \bibinfo {author} {\bibfnamefont {M.}~\bibnamefont {Hirohama}},\ }\bibfield  {title} {\bibinfo {title} {{The Implementation Of Postgres}},\ }\href {https://doi.org/10.1109/69.50912} {\bibfield  {journal} {\bibinfo  {journal} {Knowledge and Data Engineering, IEEE Transactions on}\ }\textbf {\bibinfo {volume} {2}},\ \bibinfo {pages} {125 } (\bibinfo {year} {1990})}\BibitemShut {NoStop}%
\bibitem [{\citenamefont {Maslov}\ \emph {et~al.}(2005)\citenamefont {Maslov}, \citenamefont {Young}, \citenamefont {Miller},\ and\ \citenamefont {Dueck}}]{maslov2005quantum}%
  \BibitemOpen
  \bibfield  {author} {\bibinfo {author} {\bibfnamefont {D.}~\bibnamefont {Maslov}}, \bibinfo {author} {\bibfnamefont {C.}~\bibnamefont {Young}}, \bibinfo {author} {\bibfnamefont {D.~M.}\ \bibnamefont {Miller}},\ and\ \bibinfo {author} {\bibfnamefont {G.~W.}\ \bibnamefont {Dueck}},\ }\bibfield  {title} {\bibinfo {title} {{Quantum circuit simplification using templates}},\ }in\ \href@noop {} {\emph {\bibinfo {booktitle} {Design, Automation and Test in Europe}}}\ (\bibinfo {organization} {IEEE},\ \bibinfo {year} {2005})\ pp.\ \bibinfo {pages} {1208--1213}\BibitemShut {NoStop}%
\bibitem [{\citenamefont {Paler}\ and\ \citenamefont {Basmadjian}(2022)}]{paler2022energy}%
  \BibitemOpen
  \bibfield  {author} {\bibinfo {author} {\bibfnamefont {A.}~\bibnamefont {Paler}}\ and\ \bibinfo {author} {\bibfnamefont {R.}~\bibnamefont {Basmadjian}},\ }\bibfield  {title} {\bibinfo {title} {{Energy cost of quantum circuit optimisation: Predicting that optimising Shor’s algorithm circuit uses 1 GWh}},\ }\href@noop {} {\bibfield  {journal} {\bibinfo  {journal} {ACM Transactions on Quantum Computing}\ }\textbf {\bibinfo {volume} {3}},\ \bibinfo {pages} {1} (\bibinfo {year} {2022})}\BibitemShut {NoStop}%
\bibitem [{\citenamefont {Saadatmand}\ \emph {et~al.}(2024)\citenamefont {Saadatmand}, \citenamefont {Wilson}, \citenamefont {Field}, \citenamefont {Krishnan~Vijayan}, \citenamefont {Le}, \citenamefont {Ruh}, \citenamefont {Singh~Maan}, \citenamefont {Moflic}, \citenamefont {Caesura}, \citenamefont {Paler} \emph {et~al.}}]{saadatmand2024fault}%
  \BibitemOpen
  \bibfield  {author} {\bibinfo {author} {\bibfnamefont {S.}~\bibnamefont {Saadatmand}}, \bibinfo {author} {\bibfnamefont {T.~L.}\ \bibnamefont {Wilson}}, \bibinfo {author} {\bibfnamefont {M.}~\bibnamefont {Field}}, \bibinfo {author} {\bibfnamefont {M.}~\bibnamefont {Krishnan~Vijayan}}, \bibinfo {author} {\bibfnamefont {T.~P.}\ \bibnamefont {Le}}, \bibinfo {author} {\bibfnamefont {J.}~\bibnamefont {Ruh}}, \bibinfo {author} {\bibfnamefont {A.}~\bibnamefont {Singh~Maan}}, \bibinfo {author} {\bibfnamefont {I.}~\bibnamefont {Moflic}}, \bibinfo {author} {\bibfnamefont {A.}~\bibnamefont {Caesura}}, \bibinfo {author} {\bibfnamefont {A.}~\bibnamefont {Paler}}, \emph {et~al.},\ }\bibfield  {title} {\bibinfo {title} {Superconducting qubits at the utility scale: the potential and limitations of modularity},\ }\href@noop {} {\bibfield  {journal} {\bibinfo  {journal} {arXiv e-prints}\ ,\ \bibinfo {pages} {arXiv}} (\bibinfo {year} {2024})}\BibitemShut {NoStop}%
\bibitem [{\citenamefont {Baker}\ \emph {et~al.}(2020)\citenamefont {Baker}, \citenamefont {Duckering}, \citenamefont {Hoover},\ and\ \citenamefont {Chong}}]{baker2020time}%
  \BibitemOpen
  \bibfield  {author} {\bibinfo {author} {\bibfnamefont {J.~M.}\ \bibnamefont {Baker}}, \bibinfo {author} {\bibfnamefont {C.}~\bibnamefont {Duckering}}, \bibinfo {author} {\bibfnamefont {A.}~\bibnamefont {Hoover}},\ and\ \bibinfo {author} {\bibfnamefont {F.~T.}\ \bibnamefont {Chong}},\ }\bibfield  {title} {\bibinfo {title} {Time-sliced quantum circuit partitioning for modular architectures},\ }in\ \href@noop {} {\emph {\bibinfo {booktitle} {Proceedings of the 17th ACM International Conference on Computing Frontiers}}}\ (\bibinfo {year} {2020})\ pp.\ \bibinfo {pages} {98--107}\BibitemShut {NoStop}%
\bibitem [{\citenamefont {Brandhofer}\ \emph {et~al.}(2023)\citenamefont {Brandhofer}, \citenamefont {Polian},\ and\ \citenamefont {Krsulich}}]{brandhofer2023optimal}%
  \BibitemOpen
  \bibfield  {author} {\bibinfo {author} {\bibfnamefont {S.}~\bibnamefont {Brandhofer}}, \bibinfo {author} {\bibfnamefont {I.}~\bibnamefont {Polian}},\ and\ \bibinfo {author} {\bibfnamefont {K.}~\bibnamefont {Krsulich}},\ }\bibfield  {title} {\bibinfo {title} {Optimal partitioning of quantum circuits using gate cuts and wire cuts},\ }\href@noop {} {\bibfield  {journal} {\bibinfo  {journal} {IEEE Transactions on Quantum Engineering}\ }\textbf {\bibinfo {volume} {5}},\ \bibinfo {pages} {1} (\bibinfo {year} {2023})}\BibitemShut {NoStop}%
\bibitem [{Note1()}]{Note1}%
  \BibitemOpen
  \bibinfo {note} {\protect \url {https://github.com/ioanamoflic/pandora}}\BibitemShut {NoStop}%
\bibitem [{\citenamefont {Lewis}\ \emph {et~al.}(2023)\citenamefont {Lewis}, \citenamefont {Soudjani},\ and\ \citenamefont {Zuliani}}]{lewis2023formal}%
  \BibitemOpen
  \bibfield  {author} {\bibinfo {author} {\bibfnamefont {M.}~\bibnamefont {Lewis}}, \bibinfo {author} {\bibfnamefont {S.}~\bibnamefont {Soudjani}},\ and\ \bibinfo {author} {\bibfnamefont {P.}~\bibnamefont {Zuliani}},\ }\bibfield  {title} {\bibinfo {title} {Formal verification of quantum programs: Theory, tools, and challenges},\ }\href@noop {} {\bibfield  {journal} {\bibinfo  {journal} {ACM Transactions on Quantum Computing}\ }\textbf {\bibinfo {volume} {5}},\ \bibinfo {pages} {1} (\bibinfo {year} {2023})}\BibitemShut {NoStop}%
\bibitem [{\citenamefont {Yamashita}\ and\ \citenamefont {Markov}(2010)}]{yamashita2010fast}%
  \BibitemOpen
  \bibfield  {author} {\bibinfo {author} {\bibfnamefont {S.}~\bibnamefont {Yamashita}}\ and\ \bibinfo {author} {\bibfnamefont {I.~L.}\ \bibnamefont {Markov}},\ }\bibfield  {title} {\bibinfo {title} {Fast equivalence-checking for quantum circuits},\ }in\ \href@noop {} {\emph {\bibinfo {booktitle} {2010 IEEE/ACM International Symposium on Nanoscale Architectures}}}\ (\bibinfo {organization} {IEEE},\ \bibinfo {year} {2010})\ pp.\ \bibinfo {pages} {23--28}\BibitemShut {NoStop}%
\bibitem [{\citenamefont {Burgholzer}\ and\ \citenamefont {Wille}(2020)}]{burgholzer2020advanced}%
  \BibitemOpen
  \bibfield  {author} {\bibinfo {author} {\bibfnamefont {L.}~\bibnamefont {Burgholzer}}\ and\ \bibinfo {author} {\bibfnamefont {R.}~\bibnamefont {Wille}},\ }\bibfield  {title} {\bibinfo {title} {Advanced equivalence checking for quantum circuits},\ }\href@noop {} {\bibfield  {journal} {\bibinfo  {journal} {IEEE Transactions on Computer-Aided Design of Integrated Circuits and Systems}\ }\textbf {\bibinfo {volume} {40}},\ \bibinfo {pages} {1810} (\bibinfo {year} {2020})}\BibitemShut {NoStop}%
\bibitem [{\citenamefont {van~de Wetering}(2020)}]{van2020zx}%
  \BibitemOpen
  \bibfield  {author} {\bibinfo {author} {\bibfnamefont {J.}~\bibnamefont {van~de Wetering}},\ }\bibfield  {title} {\bibinfo {title} {Zx-calculus for the working quantum computer scientist},\ }\href@noop {} {\bibfield  {journal} {\bibinfo  {journal} {arXiv preprint arXiv:2012.13966}\ } (\bibinfo {year} {2020})}\BibitemShut {NoStop}%
\bibitem [{\citenamefont {Elliott}\ \emph {et~al.}(2008)\citenamefont {Elliott}, \citenamefont {Eastin},\ and\ \citenamefont {Caves}}]{elliott2008graphical}%
  \BibitemOpen
  \bibfield  {author} {\bibinfo {author} {\bibfnamefont {M.~B.}\ \bibnamefont {Elliott}}, \bibinfo {author} {\bibfnamefont {B.}~\bibnamefont {Eastin}},\ and\ \bibinfo {author} {\bibfnamefont {C.~M.}\ \bibnamefont {Caves}},\ }\bibfield  {title} {\bibinfo {title} {Graphical description of the action of clifford operators on stabilizer states},\ }\href@noop {} {\bibfield  {journal} {\bibinfo  {journal} {Physical Review A—Atomic, Molecular, and Optical Physics}\ }\textbf {\bibinfo {volume} {77}},\ \bibinfo {pages} {042307} (\bibinfo {year} {2008})}\BibitemShut {NoStop}%
\bibitem [{\citenamefont {Paler}\ and\ \citenamefont {Devitt}(2018)}]{paler2018specification}%
  \BibitemOpen
  \bibfield  {author} {\bibinfo {author} {\bibfnamefont {A.}~\bibnamefont {Paler}}\ and\ \bibinfo {author} {\bibfnamefont {S.~J.}\ \bibnamefont {Devitt}},\ }\bibfield  {title} {\bibinfo {title} {Specification format and a verification method of fault-tolerant quantum circuits},\ }\href@noop {} {\bibfield  {journal} {\bibinfo  {journal} {Physical Review A}\ }\textbf {\bibinfo {volume} {98}},\ \bibinfo {pages} {022302} (\bibinfo {year} {2018})}\BibitemShut {NoStop}%
\bibitem [{\citenamefont {Ullman}\ and\ \citenamefont {Widom}(1997)}]{DBMS}%
  \BibitemOpen
  \bibfield  {author} {\bibinfo {author} {\bibfnamefont {J.~D.}\ \bibnamefont {Ullman}}\ and\ \bibinfo {author} {\bibfnamefont {J.}~\bibnamefont {Widom}},\ }\href@noop {} {\emph {\bibinfo {title} {A first course in database systems}}}\ (\bibinfo  {publisher} {Prentice-Hall, Inc.},\ \bibinfo {address} {USA},\ \bibinfo {year} {1997})\BibitemShut {NoStop}%
\bibitem [{\citenamefont {Cross}\ \emph {et~al.}(2022)\citenamefont {Cross}, \citenamefont {Javadi-Abhari}, \citenamefont {Alexander}, \citenamefont {De~Beaudrap}, \citenamefont {Bishop}, \citenamefont {Heidel}, \citenamefont {Ryan}, \citenamefont {Sivarajah}, \citenamefont {Smolin}, \citenamefont {Gambetta} \emph {et~al.}}]{cross2022openqasm}%
  \BibitemOpen
  \bibfield  {author} {\bibinfo {author} {\bibfnamefont {A.}~\bibnamefont {Cross}}, \bibinfo {author} {\bibfnamefont {A.}~\bibnamefont {Javadi-Abhari}}, \bibinfo {author} {\bibfnamefont {T.}~\bibnamefont {Alexander}}, \bibinfo {author} {\bibfnamefont {N.}~\bibnamefont {De~Beaudrap}}, \bibinfo {author} {\bibfnamefont {L.~S.}\ \bibnamefont {Bishop}}, \bibinfo {author} {\bibfnamefont {S.}~\bibnamefont {Heidel}}, \bibinfo {author} {\bibfnamefont {C.~A.}\ \bibnamefont {Ryan}}, \bibinfo {author} {\bibfnamefont {P.}~\bibnamefont {Sivarajah}}, \bibinfo {author} {\bibfnamefont {J.}~\bibnamefont {Smolin}}, \bibinfo {author} {\bibfnamefont {J.~M.}\ \bibnamefont {Gambetta}}, \emph {et~al.},\ }\bibfield  {title} {\bibinfo {title} {{OpenQASM 3: A broader and deeper quantum assembly language}},\ }\href@noop {} {\bibfield  {journal} {\bibinfo  {journal} {ACM Transactions on Quantum Computing}\ }\textbf {\bibinfo {volume} {3}},\ \bibinfo {pages} {1} (\bibinfo {year} {2022})}\BibitemShut {NoStop}%
\bibitem [{\citenamefont {Harrigan}\ \emph {et~al.}(2024)\citenamefont {Harrigan}, \citenamefont {Khattar}, \citenamefont {Yuan}, \citenamefont {Peduri}, \citenamefont {Yosri}, \citenamefont {Malone}, \citenamefont {Babbush},\ and\ \citenamefont {Rubin}}]{harrigan2024qualtran}%
  \BibitemOpen
  \bibfield  {author} {\bibinfo {author} {\bibfnamefont {M.~P.}\ \bibnamefont {Harrigan}}, \bibinfo {author} {\bibfnamefont {T.}~\bibnamefont {Khattar}}, \bibinfo {author} {\bibfnamefont {C.}~\bibnamefont {Yuan}}, \bibinfo {author} {\bibfnamefont {A.}~\bibnamefont {Peduri}}, \bibinfo {author} {\bibfnamefont {N.}~\bibnamefont {Yosri}}, \bibinfo {author} {\bibfnamefont {F.~D.}\ \bibnamefont {Malone}}, \bibinfo {author} {\bibfnamefont {R.}~\bibnamefont {Babbush}},\ and\ \bibinfo {author} {\bibfnamefont {N.~C.}\ \bibnamefont {Rubin}},\ }\href {https://doi.org/10.48550/arXiv.2409.04643} {\bibinfo {title} {Expressing and analyzing quantum algorithms with qualtran}} (\bibinfo {year} {2024}),\ \Eprint {https://arxiv.org/abs/2409.04643} {arXiv:2409.04643 [quant-ph]} \BibitemShut {NoStop}%
\bibitem [{Note2()}]{Note2}%
  \BibitemOpen
  \bibinfo {note} {\protect \url {https://github.com/quantumlib/Cirq}}\BibitemShut {NoStop}%
\bibitem [{\citenamefont {K.}\ \emph {et~al.}(2025)\citenamefont {K.}, \citenamefont {J.}, \citenamefont {K.}, \citenamefont {B.}, \citenamefont {P.}, \citenamefont {R.}, \citenamefont {A.}, \citenamefont {R.}, \citenamefont {J.},\ and\ \citenamefont {J.}}]{pyliqtr}%
  \BibitemOpen
  \bibfield  {author} {\bibinfo {author} {\bibfnamefont {O.}~\bibnamefont {K.}}, \bibinfo {author} {\bibfnamefont {E.}~\bibnamefont {J.}}, \bibinfo {author} {\bibfnamefont {M.}~\bibnamefont {K.}}, \bibinfo {author} {\bibfnamefont {R.}~\bibnamefont {B.}}, \bibinfo {author} {\bibfnamefont {K.}~\bibnamefont {P.}}, \bibinfo {author} {\bibfnamefont {N.}~\bibnamefont {R.}}, \bibinfo {author} {\bibfnamefont {K.}~\bibnamefont {A.}}, \bibinfo {author} {\bibfnamefont {R.}~\bibnamefont {R.}}, \bibinfo {author} {\bibfnamefont {B.}~\bibnamefont {J.}},\ and\ \bibinfo {author} {\bibfnamefont {B.}~\bibnamefont {J.}},\ }\href {https://doi.org/10.5281/zenodo.14719561} {\bibinfo {title} {pyliqtr (v1.3.6)}} (\bibinfo {year} {2025})\BibitemShut {NoStop}%
\bibitem [{Note3()}]{Note3}%
  \BibitemOpen
  \bibinfo {note} {Storing quantum circuits on disk can be effectively used as a way to cache circuit decompositions for later use}\BibitemShut {NoStop}%
\bibitem [{Note4()}]{Note4}%
  \BibitemOpen
  \bibinfo {note} {The Pandora DAG format is resource-efficient in terms of storage requirements, as every billion of gates takes about 100GB of disk space.}\BibitemShut {Stop}%
\end{thebibliography}%

\section{Appendix}

\subsection{Storing Quantum Circuits in a Database}
\label{sec:database}
Relational Database Management Systems (RDBMS) organize data into tables composed of rows and columns~\cite{DBMS}. Each row (also called entry or tuple) of a table is identified by a unique id and the columns give the features of each uniquely identified row. The database can be queried to obtain information about the data and queries are generally written in the Structured Query Language (SQL). SQL is the standard programming language for interacting with RDBMS, and allows the database user to add, update, or delete rows of data.

One of the advantages of RDBMS is that it allows for a large number of queries to be executed almost in parallel by multiple users, without compromising the integrity of the data. This is achieved by communicating with the database via independent transactions, which obey the widely known ACID properties (atomicity, consistency, isolation, durability). These exact properties ensure data is not corrupted during read and writes when transaction concurrency is involved. The strictness of protection against data corruption is given by the transaction isolation levels available in most database systems. 

We store quantum circuits as tables and apply template rewrite rules in the form of database queries. A quantum circuit is stored in a tabular format in a way that allows for \emph{perfect reconstruction}(lossless representation) of the circuit. Each gate of the quantum circuit is a separate table row. When applied to circuits as database tables, the rewrite functions are: 
\begin{itemize}
    \item searching for table rows corresponding to a set of criteria (e.g. gate type);
    \item deleting or inserting rows into the table;
    \item updating rows.
\end{itemize}

\begin{table*}[t]
\centering
 \begin{tabular}{c |c |c |c |c |c |c |c |c |c| c} 
  ID & PREV\_Q1 & PREV\_Q2 & PREV\_Q3 & TYPE & PRM & SWITCH & NEXT\_Q1 & NEXT\_Q2 & NEXT\_Q3 & LABEL\\ [0.5ex] \hline
 1 & null & null & null & CNOT & 0 & true & 2 & 3 & null & hyp\_circ\\ 
 2 & 1 & null & null & H & 0 & false & null & null & null &hyp\_circ\\ 
 3 & 1 & null & null & H & 0 & false & null & null & null &hyp\_circ\\ 
 \end{tabular}
\caption{A hypothetical three qubit circuit (Fig.~\ref{fig:example}) composed of three gates. The gates of the circuit are fully described by the columns \textit{ID}, \textit{PREV\_Q1}, \textit{PREV\_Q2}, \textit{PREV\_Q3},  \textit{GATE\_TYPE},  \textit{PARAM.}, \textit{SWITCH}, \textit{NEXT\_Q1}, \textit{NEXT\_Q2}, \textit{NEXT\_Q3}, \textit{LABEL}. Each gate in the circuit is uniquely identified by its \textit{ID} value. Each circuit in the table is uniquely identified by its LABEL value.}
\label{tab:example}
\end{table*}

\begin{figure}[!h]
    \centering
    \includegraphics[width=0.15\columnwidth]{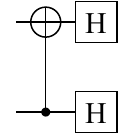}
    \caption{The circuit from Table~\ref{tab:example}.}
    \label{fig:example}
\end{figure}

\begin{figure}[!t]
    \centering
    \includegraphics[width=0.8\columnwidth]{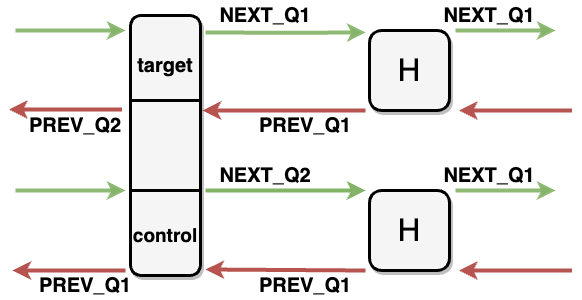}
    \caption{The diagram that represents the circuit stored in Table \ref{tab:example} has a doubly linked list structure. Each gate points to its neighbouring gates for each qubit that it acts on. The single qubit Hadamard gate only has one previous gate and one following gate. On the other hand, the CNOT gate will point to at most two previous gates and at most two following gates.}
    \label{fig:linked}
\end{figure}

Our database model of a quantum circuit is similar to a graph representation: we store gates similarly to vertices and use \emph{pointers} to the neighbouring gates/vertices. The graph representation of the circuit is efficient as it also allows for seamless logical synthesis of gates. Complex gates can easily be decomposed, for example a Toffoli gate into Clifford+T, by replacing a database row with the corresponding decomposition.

Currently, our RDBMS quantum circuit model supports single, two and three qubit gates. The notion of qubit is lost in this database circuit representation: for every database row that represents a gate, we only record a \emph{pointer} of the immediately neighbouring gate which acts on the same qubits (Fig.~\ref{fig:linked}). The values of the \texttt{PREV\_Qx} and \texttt{NEXT\_Qx} (with \texttt{x} in \{1,2,3\}) columns are filled according to a set of predefined rules:

\begin{enumerate}
    \item Each single qubit gate acts on qubit \texttt{Q1}. The only columns that store a non-null value for the neighbours of the gate are \texttt{PREV\_Q1} and \texttt{NEXT\_Q1}. 
    
    \item Each two-qubit gate acts only on two qubits: \texttt{Q1} and \texttt{Q2}. The only columns that store a non-null value for the neighbours of the gate are \texttt{PREV\_Q1},  \texttt{PREV\_Q2},  \texttt{NEXT\_Q1} and \texttt{NEXT\_Q2}. For two-qubit gates, the convention is that the control always acts on \texttt{Q1} and the target on \texttt{Q2}.
    
    \item The only three-qubit gate considered in this representation is the Toffoli gate, where \texttt{Q1} is the first control, \texttt{Q2} is the second control and \texttt{Q3} is the target. All fields that represent a previous or next value should be non-null.
\end{enumerate}

We use two special types of gates, namely \textit{Initial (In)} and \textit{Final (Out)} gates. These resemble single-qubit gates, and can be used similarly to barriers~\cite{cross2022openqasm} to avoid qubit permutations during reconstruction and ensure the correctness of the rewrite procedure. The \textit{In} gate stores a non-null value for \texttt{NEXT\_Q1}, while \texttt{PREV\_Q1} will always be null. Similarly, the \textit{Out} gate will only store a non-null value for \texttt{PREV\_Q1}.

\begin{figure}[!t]
    \centering
    \includegraphics[width=1\columnwidth]{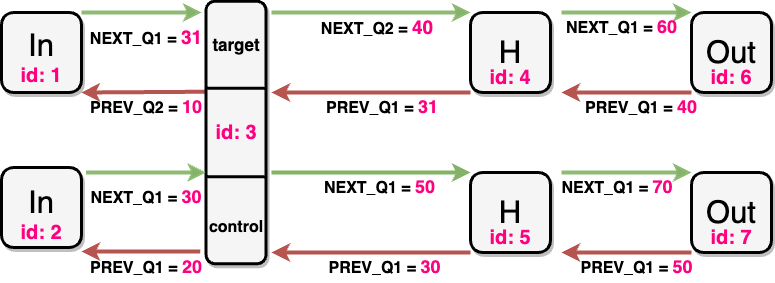}
    \caption{Modifying Fig.~\ref{fig:linked} in order to force it to obey the previously mentioned rules results in a graph-like structure with four more nodes. Each node of the graph is a gate uniquely identified by an ID. For example, the CNOT from Fig.~\ref{fig:example} is encoded into a single node with four incoming edges and four outgoing ones. The role of these edges is to create a doubly-linked list-like data structure which can replace the notion of qubit or time-slice present in the diagrammatic representation of quantum circuits.}
    \label{fig:linked_final}
\end{figure}

The \texttt{PREV} and \texttt{NEXT} columns are used as helpers for database operations (see following sections) which apply templates such as the ones from Fig.~\ref{fig:rules}. Fig.~\ref{fig:linked_final} illustrates the construction and use of the \texttt{PREV} or \texttt{NEXT} values. Therein, the value stored in each \texttt{PREV} or \texttt{NEXT} column is the concatenation of the neighbouring gate \texttt{ID} and one of three postfixes from the set $\{0, 1, 2\}$. The postfixes have the following meaning: a) the value $0$ signals that one of \texttt{PREV} or \texttt{NEXT} is shared with a control of a two/three qubit gate or with a single qubit gate; b) the value $1$ stands for sharing the qubit with the target of a neighbouring two or three qubit gate; c) value $2$ stands for the second control of a neighbouring Toffoli gate. Fig.~\ref{fig:linked_final} illustrates the construction and use of the \texttt{PREV} or \texttt{NEXT} values.

The advantage of our graph-like, doubly-linked-list quantum circuit representation is that deleting and inserting gates comes at a reduced computational overhead. The rewrite rules that delete or insert a row have to update only the \texttt{PREV} and \texttt{NEXT} columns accordingly. 

Nevertheless, during reconstruction, each input gate will be assigned a qubit number which will be passed to its children during the level order traversal of the tree. A perfect reconstruction of the circuit is achievable without necessarily preserving a certain order of the database entries.

\subsection{Implementation of the Partitioning Method}

The partitioning method from Section~\ref{sec:part} is:
\begin{enumerate}
    \item We begin by generating the topologically-ordered edge list of $(V_i, V_{i+1})$ and cache the list into Pandora, where each $V_k$ is a node of the DAG. Each element of the list is identified with a directed edge from node $V_i$ to node $V_{i+1}$. This step is performed only once;
    
    \item We continue by applying the \texttt{union} and \texttt{find} operations and grow node partitions as long as no bounds are exceeded (e.g. the T-count of the partition is lower or equal than the maximum allowed);
    
    \item We return the generated partitions.
\end{enumerate}

We perform partitioning in a streamed fashion. Separately in Pandora, we pre-generate the list of edges which are consumed by the partitioning algorithm. Practically, we perform the following steps:
\begin{enumerate}
    \item An arbitrary circuit is loaded from \texttt{Qualtran}/\texttt{pyLIQTR} into Pandora;
    \item The Pandora representation is transpiled to elementary gates (e.g. Clifford+T/Clifford+Rz);
    \item We extract the DAG representation of the resulting circuit and cache it;
    \item We select a parameter range for numerically investigating the benefits of partitioning (e.g. a partition should have between 10 and 100 T gates);
    \item We grow the partitions;
    \item Finally, we analyze the resulting partitions for structural similarities.
\end{enumerate}

\begin{figure*}[!t]
    \centering
    \includegraphics[width=0.9\textwidth]{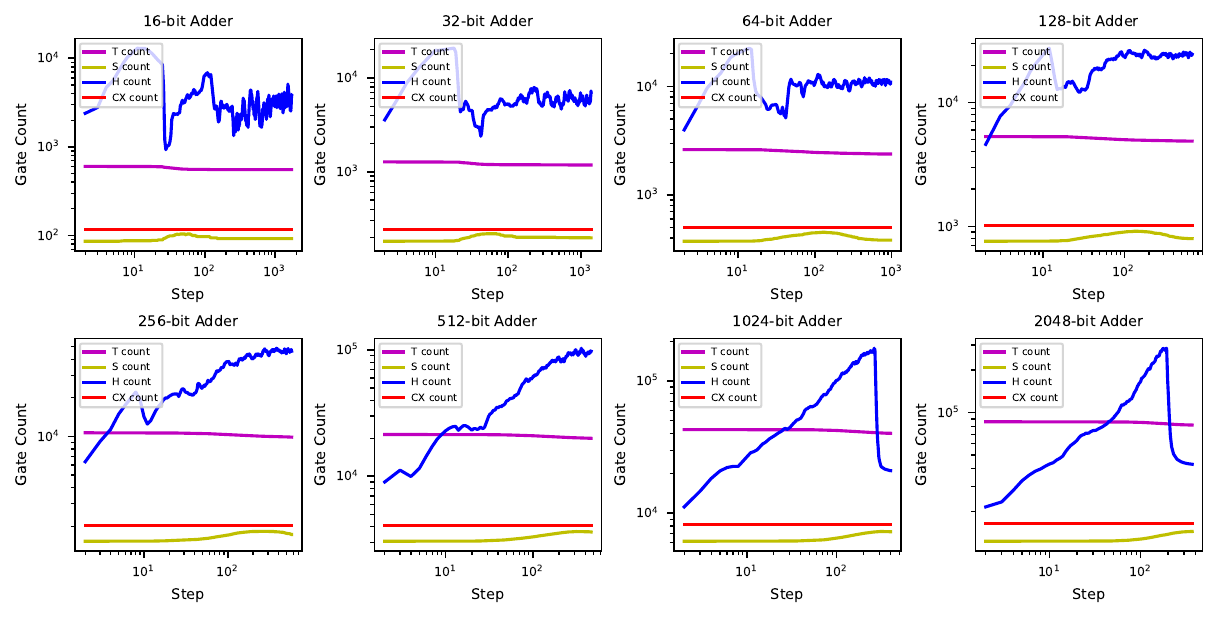}
    \caption{The results of benchmarking optimisation on Adders 16-2048 from~\cite{nam2018automated} with a timeout of 600s. Each circuit was optimised in a parallel fashion by letting 24 threads interact for a limited time of 10 minutes. The threads were each assigned one of the templates from Fig.~\ref{fig:rules}. The intuition behind the T count reduction gained by analysing Fig.~\ref{fig:adder_bench} is that even with a completely blind heuristic, a T count reduction is achievable when both the number of Hadamard and S gates increase. This is expected as using REVERSE and HADAMARD will most likely allow for T and S gates to be commuted through the circuit and further canceled.}
    \label{fig:adder_bench}
\end{figure*}

\subsection{\texttt{Qualtran} and \texttt{pyLIQTR} interoperability}

Pandora can ingest algorithms from both Google \texttt{Qualtran} and \texttt{pyLIQTR} and relies on their ability to decompose high-level algorithms down to designated gate sets such as Clifford+T or Clifford+Rz. The decomposition itself is achieved either via Cirq or \texttt{pyLIQTR}.

\texttt{Qualtran}~\cite{harrigan2024qualtran} (Quantum Algorithms Translator) is an open source tool for analysing and estimating the resources of quantum algorithms. \texttt{Qualtran} allows its user to define different algorithms as structures made out of Bloqs, which are typically building blocks of high-level algorithms. Bloqs usually come in the form of Adders, Multipliers, QROM, Quantum Phase Estimation (QPE), etc. 

\texttt{Qualtran} allows logical algorithms to be decomposed to Clifford+T by using hybrid Cirq and \texttt{Qualtran} functionalities. \texttt{Qualtran} Bloqs have a recursive-like, hierarchical pattern: abstract composite Bloqs are composed of other composite Bloqs which can further be decomposed until atomic Bloqs are encountered. \texttt{Qualtran} and Cirq are syntactically very similar, and one feature of \texttt{Qualtran} is Cirq \footnote{\url{https://github.com/quantumlib/Cirq}} interoperability, which means \texttt{Qualtran} Bloqs can be converted to Cirq circuits and vice-versa. 

\texttt{pyLIQTR}~\cite{pyliqtr} is an additional tool that can be used to obtain resource estimates of quantum circuits built out of quantum algorithms defined in \texttt{Qualtran}.

The definition of Bloq decompositions in \texttt{Qualtran} allows one to decompose high-level circuits until a specific criteria is met. One of these criteria could be that the final circuit consists solely of gates from a designated gate set. For very large circuits, decomposition times proved to be a major performance bottleneck. Although Pandora has the capacity to store circuits with billions of gates, the decomposition times can potentially make it very difficult for circuits to be processed quickly.

In order to speed up the decomposition, we switched from Cirq's decomposition method to \texttt{pyLIQTR}'s decomposition, which is significantly faster because it is based on Python generators. For added speed, we use the caching capabilities of Pandora~\footnote{storing quantum circuits on disk can be effectively used as a way to cache circuit decompositions for later use}, and implemented a streamed decomposition pipeline in Pandora. This pipeline allows us to process very large circuits in batches. The batching of the decomposition is important for reducing the memory footprint; only elements of a single batch occupy memory at a given time.

The streamed decomposition method returns a batch of Clifford+T gates, and the batch is translated to the Pandora DAG format~\footnote{The Pandora DAG format is resource-efficient in terms of storage requirements, as every billion of gates takes about 100GB of disk space.} and then inserted into the database. The decomposition method runs concurrently to the Pandora insertion, and this achieves performance speedups.  

As a result of these performance improvements and multi-threading, we can compile circuits such as utility-scale $100\times100$ Fermi-Hubbard instances in a matter of a few hours. 

\end{document}